\PassOptionsToPackage{lstlisting,prologue,dvipsnames}{xcolor}

\documentclass[conference]{IEEEtran}
\IEEEoverridecommandlockouts

\usepackage[dvipsnames]{xcolor} 
\definecolor{LightGray}{rgb}{0.97,0.97,0.97}
\usepackage{cite}
\usepackage{amsmath,amssymb,amsfonts}
\usepackage{algorithmic}
\usepackage{graphicx}
\usepackage{textcomp}
\usepackage{xcolor}
\usepackage{url}
\usepackage[final]{listings}
\usepackage{listing-td}
\usepackage{amsmath}
\usepackage{empheq}
\usepackage{multirow}

\usepackage{lipsum}
\usepackage{graphics}
\usepackage{array}
\usepackage{listing-td}
\usepackage{listing-sparql}
\usepackage{hyperref}
\usepackage{listings}
\usepackage{amsmath}
\usepackage{lipsum}
\usepackage{graphics}
\usepackage{array}
\usepackage{mdframed}

\def\BibTeX{{\rm B\kern-.05em{\sc i\kern-.025em b}\kern-.08em
    T\kern-.1667em\lower.7ex\hbox{E}\kern-.125emX}}

\begin{document}

\title{Towards Hypermedia Environments for \\Adaptive Coordination in Industrial Automation}

\author{\IEEEauthorblockN{1\textsuperscript{st} Ganesh Ramanathan}
\IEEEauthorblockA{\textit{Siemens AG}\\
Zug, Switzerland \\
ganesh.ramanathan@siemens.com}
\and
\IEEEauthorblockN{2\textsuperscript{nd} Simon Mayer}
\IEEEauthorblockA{\textit{University of St. Gallen}\\
Switzerland \\
simon.mayer@unisg.ch}
\and
\IEEEauthorblockN{3\textsuperscript{rd} Andrei Ciortea}
\IEEEauthorblockA{\textit{University of St. Gallen}\\
Switzerland \\
andrei.ciortea@unisg.ch}
}
\maketitle

\begin{abstract}
Electromechanical systems manage physical processes through a network of inter-connected components.
Today, programming the interactions required for coordinating these components is largely a manual process. This process is time-consuming and requires manual adaptation when system features change.
To overcome this issue, we use autonomous software agents that process semantic descriptions of the system to determine coordination requirements and constraints; on this basis, they then interact with one another to control the system in a decentralized and coordinated manner.
%Though Multi-Agent Systems (MAS) have been proposed in the past to tackle this challenge, a practical adoption in industrial automation is yet to be seen.
%Our research aims to show that this can be overcome if autonomous software agents can use semantic descriptions of the system to determine coordination needs and constraints.
Our core insight is that coordination requirements between individual components are, ultimately, largely due to underlying physical interdependencies between the components, which can be (and, in many cases, already are) semantically modeled in automation projects.
%This is supported as large parts of the coordination requirements between individual components are due to underlying physical inter-dependencies of the components, which can be (and, in many cases, already are) semantically modeled in automation projects.
%Further, we show that the concept of hypermedia MAS supports the vision of achieving coordination in dynamic environments by avoiding tight coupling. 
Agents then use hypermedia to discover, at run time, the plans and protocols required for enacting the coordination.
A key novelty of our approach is the use of hypermedia-driven interaction: it reduces coupling in the system and enables its run-time adaptation as features change.
%We demonstrate the potential benefits of our approach using a real-life scenario of a heating system in a building, which shows promising results and highlights challenges for further research.

\end{abstract}



\maketitle
\section{Introduction}
Systems in installations like chemical processing plants, factories, and buildings are composed of electro-mechanical components that progressively modify substances and manipulate forces and energy forms towards desired states.
Ultimately, each of the components involved in such processes manages a set of physical mechanisms controlled by a software program. We use the term \textit{agent} to refer to such programs that are responsible for controlling the functioning of one or more components.
However, the functioning of the components is often also dependent on their peers, which are located up- and downstream in the process.
Coordination between the agents is hence required to orchestrate the physical mechanisms and ensure safety conditions at the startup, operation, and shutdown of the process.

In deployed systems, coordination is typically implemented in one of two broad architectural schemes: (1) through the use of hierarchical supervisory control, or (2) by distributing the responsibility amongst the participants (e.g., in complex large-scale deployments).
For both cases, the \textit{coordination plans} (i.e., when and how the agents interact) are synthesised by a combination of top-down design decomposition and bottom-up integration of component features. In current systems, these plans are \emph{hard-coded in the agents} and any change in the system features requires re-engineering~\cite{rasmusson1997decentralized,chokshi2007distributed}.
This hard coding affects especially distributed process control in scenarios that face reconfiguration during the system's lifetime (e.g., flexible manufacturing or changing usage patterns in building control systems)~\cite{ciortea2018repurposing,rasmusson1997decentralized,chokshi2007distributed}.

Though several researchers have advocated the use of Multi-agent Systems (MAS) for industrial automation that have to deal with dynamic environments~\cite{ciortea2018repurposing,wang2017coordination,teran2017integration,maturana2004distributed}, the question of how agents would determine their \textit{coordination responsibilities} so that they discover and interact with collaborators is still largely under-explored.
Our past research has demonstrated that current engineering ontologies based on Semantic Web technologies can be integrated to assist agents in choosing control strategies for local functions~\cite{ramanathan2023reasoning,ramanathan2023synthesizing,case2024}. In this work, we show that agents can also use such ontologies to infer coordination responsibilities.
But in contrast to local controls, coordination requires the agents to locate and interact with peers. To enable this to happen in a loose-coupled and interoperable manner, we motivate the adoption of hypermedia-driven interaction where agents utilize a Web-based environment to publish their profiles, navigate using semantic links, and discover affordances to set up and enact the coordination dynamically.

\section{Related Work}

Engineering of coordination is based on the knowledge of system requirements, technical design, models of physical processes, and the control strategies~\cite{sd_nielsen2015systems,sd_van2013ontology,sd_yang2019ontology}.
One interoperable and extensible way to represent these descriptions is through ontologies such as BRICK~\cite{sd_balaji2016brick} and OntoCAPE~\cite{morbach2009ontocape} for topologies, PhySys~\cite{borst1995physsys} and Elementary~\cite{ramanathan2023reasoning} for physical processes, CTRLOnt~\cite{preisig2021ontology} for control programs, and GORE~\cite{negri2017towards} for requirements.
Our work~\cite{case2024} has shown that these ontologies can be \emph{integrated} using high-level concepts and that, in this way, control agents can know their context of operation`(cf.~\cite{ramanathan2023synthesizing}) and select suitable control programs for operation of the component.
This knowledge can be extended further to describe the physical effects of actions on a component (cf.~\cite{ramanathan2023reasoning}).
\cite{schneider2017ontology} has proposed an ontology to model finite state machines which can also be used for modeling coordination program of a supervisory controller~\cite{preisig2021ontology}.
When it comes to systems that need frequent reorganization,~\cite{ciortea2019decade,ciortea2018repurposing} have shown the benefits of placing industrial software agents in a hypermedia environment, introducing hypermedia MAS as a new class of Web-based MAS; and~\cite{vachtsevanou2023signifiers,vachtsevanou2024enabling} has shown that agents can use standardized W3C Web of Things \emph{Thing Descriptions}~\footnote{\url{https://www.w3.org/TR/wot-thing-description11/}} (WoT TDs) to describe not just their interfaces but also reveal their affordances in a way that considers the run-time situation of the agent-environment system.
We seek to combine the benefits of these individual approaches to achieve adaptive coordination in distributed process control. 

\begin{figure}[ht]
\centerline{\includegraphics[width=1.0\linewidth]{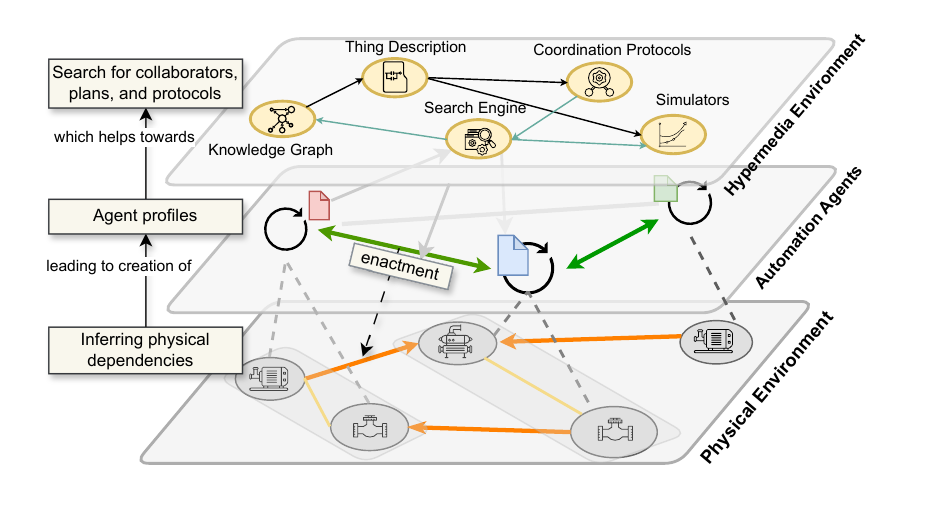}}
\caption{In our approach agents use bottom-up knowledge of physical processes to search the hypermedia environment and discover at run time means of enacting the required coordination. }
\label{fig:overview}
\end{figure}

\section{An Integration of Approaches}

In our research, we show that by being part of a hypermedia environment, agents in industrial automation are enabled to discover their collaboration partners and autonomously align themselves to fulfill coordination responsibilities. To this end, in Section~\ref{sec:infer-coordination} we show how existing machine-understandable knowledge about the system can be used to infer physical process dependencies between the components and how this can be used to establish the \textit{need for coordination} with their dependent counterparts (in Section~\ref{sec:discover-agents}).
On this basis, the enactment of coordination requires an agent to interact with others (e.g., to query and update states or invoke actions) according to a plan. However, today, the interfaces between programs are coupled using communication protocols and \emph{design-time} knowledge about information resources---again, if these interfaces change, manual re-engineering would be required. To overcome this, in Section~\ref{sec:enactment} we propose the use of semantically described \textit{coordination protocols} together with the Web of Things paradigm to achieve loose coupling of agents and interaction opportunities, and thereby run-time adaptation of the agents. 

\begin{figure}
    \centerline{\includegraphics[width=0.8\linewidth]{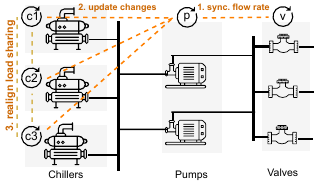}}
    \caption{A chilled water refrigeration plant where agents have to coordinate flow control and heat-extraction processes}
    \label{fig:chiller}
\end{figure}

Our work therefore connects existing research in control and process engineering, the Semantic Web, and hypermedia MAS to show that adaptive coordination of process control in large dynamic systems can be achieved.
Figure~\ref{fig:overview} provides a high-level overview, which contextualizes the individual aspects that we describe below.
Throughout this section, we use the example of a chilled water production system shown in Figure \ref{fig:chiller} as a running example~\footnote{Can be accessed here \url{https://autonomous-buildings.github.io/hmas}}. In many places, the listings in this paper have been simplified to convey the gist of the approach more easily.
%Also, Figure~\ref{fig:overview} provides a high-level overview, which contextualizes the individual aspects that we describe below.

\begin{figure}
    \centerline{\includegraphics[width=1.1\linewidth]{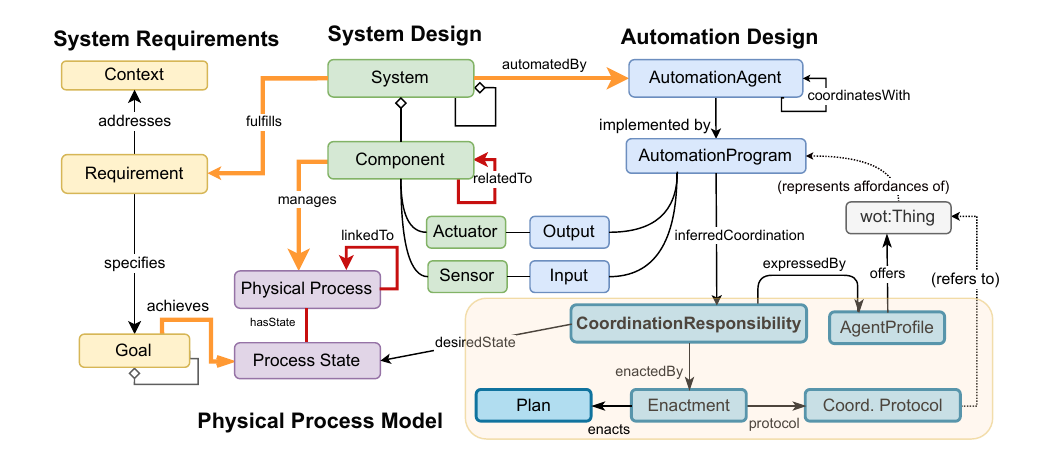}}
    \caption{We use the integrated system description to infer coordination needs and its enactment (highlighted box).}
    \label{fig:ontology}
\end{figure}

\subsection{Inferring Coordination Responsibilities}
\label{sec:infer-coordination}
Figure \ref{fig:ontology} shows the inter-connected high-level concepts in the current ontologies for representing the different aspects of system knowledge ~\cite{ramanathan2023synthesizing}.
We introduce the concept of \textit{coordination responsibility} (CR) to capture the knowledge about \textit{why} an agent needs to coordinate with others and show that this can be obtained through semantic queries on the integrated system description.
Specifically, given the process goals of the system and the set of components that an agent manages, CR is inferred in terms of \textit{process dependencies} between the components within an agent's scope and those outside it.
%If a dependent variable of a component is manipulated by an actuator, then we assume that there could be an automation agent influencing it.
%Two components are considered inter-related if the dependent process variable of one is physically connected to influence the independent variable of the other.
The inferred CR also allows the agent to list its own \textit{affordances} that may be relevant for its collaborating partners.
An affordance provided by an agent, for example, can be the expression of its current state, an action that it can be asked to take, or a goal to fulfill autonomously.
In section \ref{sec:enactment} we describe how WoT TD enables machine-understandable description of such affordances.
The agent combines these to create an \textit{agent profile} (AP). 

In the example of the chilled water system, the CR of the agent managing the pumps can be inferred using a semantic query (listed using SPARQL and the result in RDF):

\begin{lstlisting}[language=SPARQL]
SELECT ?comp ?rel ?dep WHERE{
:agent-p elem:manages ?subsystem.
?subsystem elem:hasComponent* ?comp.
?comp ?rel ?dep.
?rel a elem:processRelation.
\end{lstlisting}

Part of the results to the query, as shown below, expresses that \texttt{pump-1}, which is managed by \texttt{agent-p}, is physically dependent on the valves downstream and influences the chillers upstream:
\begin{lstlisting}[language=SPARQL]
:pump-1 a elem:Component, brick:Pump.
elem:affectedVariable :flow-rate;
elem:influences :chlr-1,chlr-2,chlr-3;
elem:influencedBy :vlv-1,vlv-2,vlv-3:
\end{lstlisting}

Therefore, \texttt{agent-p} has to coordinate with the agents responsible for the chillers and valves---but it does not know these agents upfront and has to discover them.
Therefore, the agent creates a profile that conveys its goals and affordances. In this case, \texttt{agent-p} expresses that other agents can ask it to achieve a particular flow rate or query its current state:

\begin{lstlisting}[language=SPARQL]
:profile-p a elem:AgentProfile;
elem:hasCR :cr-p-process-dependencies;
td:hasActionAffordance :change-flowrate;
td:hasPropertyAffordance :current-flowrate;
...
\end{lstlisting}
% The Agent now knows that it needs to coordinate, but it does not know with which other agents it needs to establish interactions to coordinate.
In the next step, we show how an agent uses its profile to navigate the hypermedia environment to discover other agent(s) with which it needs to establish coordination interaction.

\subsection{Discovering Coordination Partners}
\label{sec:discover-agents}
% How does the agent find collaborators based on the agent profile
% The coordination requirement now lives in the mind of the agent. 
% Simon think there are now the two "obvious" ways: direct A2A or coordination via A2E...
Traditionally, the deployment scheme of agents in the automation system would be known upfront front the design. But we are aiming to cater for a system that can be dynamically reconfigured.
For this purpose, we argue that hypermedia environments are valuable in industrial automation due to their ability to integrate diverse types of data and media formats, such as knowledge graphs, TDs, functional mock-up units for simulation, into a cohesive framework~\cite{ciortea2020autonomous}. This integration facilitates more efficient information retrieval because agents are redirected to resources in a contextualized manner.
For example, an agent may submit its profile to a registry service, which could extract the TD and publish it to a repository or query the knowledge graph to determine recipients likely to be interested in the profile.
If the profile is not of interest to a recipient, but the recipient knows a further context, it may simply redirect the profile there.
Therefore, agents can navigate complex systems more intuitively, accessing relevant information quickly and effectively. Furthermore, hypermedia environments support dynamic updates and notification of changes, which are crucial when the system's state or even its design changes at run time.

%---the concept of HMAS and why it is relevant to this research---

%Having found its collaborating partners, an agent now intiates a phase of negotiations that reveals to it the interaction protocol to use and the TDs which contain the affordances which correspond to the messages in the protocol

\subsection{Enacting the Coordination}
\label{sec:enactment}
% The agents know each other and know the coordination requirements. Now they need to figure out the actual TDs that they need to use for acting upon these requirements.
In Section \ref{sec:infer-coordination}, we have seen that an agent seeks counterparts to collaborate by advertising its profile, a part of which describes the interaction interfaces it offers to the outside world.
Complementary to this, potentially collaborating agents also expose their profiles.

We propose to use WoT TD to describe an agent's contextualized affordances, including ways to observe its state or act on it (or request it to act).
In ~\cite{case2024}, we have shown that TDs can be embedded with machine-understandable knowledge about physical mechanisms.
Based on this, the agent profiles can be semantically matched to determine if the process goals envisaged in the CR are achievable through the use of affordances exposed by the agents. 
%WoT TD was conceived for semantically describing physical components and their technical interfaces in a machine-understandable manner.
%But seen from a practical viewpoint, the (digital) affordances of a component are actually offered by some form of control or monitoring agent which manages the component.
The following example illustrates a TD exposed by \texttt{agent-c1} in response to the profile of the \texttt{agent-p}:

\begin{lstlisting}[language=td,numbers=none, caption={An example TD of an agent that manages a chiller}, label={lst:td-boiler-01}]
{ "@id": "urn:agent-c1", "@type": "hmas:Agent",
  "links": ["urn:chiller-2","urn:chiller-3"], 
  
  "properties": { "present-output": {
      "@type": "hvac:CoolingPower",
      "observes": ":variable_energy-output",
      "readOnly": true, "type": "number",
      "forms": ["href":"/states/output"]}
  },
  "actions": { "set-flowrate": {
      "@type": "hvac:FlowRate",
      "manipulates": ":variable_water-flow-rate",
      "forms": [{"href":"/actions/set-flow"}] }
  }}
\end{lstlisting}

If an agent has nothing to offer towards collaboration, it may result in an empty TD (i.e. without any affordances). In such cases, the caller is often redirected to another agent through the use of hypermedia links embedded in the TD, highlighting their crucial role in the process.
The following example query finds affordances that affect the desired physical state of the process (and therefore relevant to the CR):

\begin{lstlisting}[language=SPARQL]
SELECT ?affordance WHERE{
 :agent-p elem:hasProfile ?ap_p.
 :agent-c1 elem:hasProfile ?ap_c1.
 ?ap_p elem:desiredState ?process_state.
 ?ap_c1 elem:hasTD ?td_c1.
 ?td_c1 td:hasActionAffordance ?affordance.
 ?affordance elem:manipulates ?process_state.
}
\end{lstlisting}

Once the agents infer that affordances required for enacting the CR exist in each other's TDs, they then need to set up the coordination enactment, which comprises two parts: (1) the plan or sequence of actions to take and (2) the protocol to use for communicating the requests (for action or status update).

We see two possible ways in which this can happen: (1) the agents search for a plan and a corresponding interaction protocol created by human experts, which are made available in a library as Web resources, or (2) they use the machine-understandable knowledge of physical mechanisms and simulation models to \textit{creatively synthesize} a coordination plan (and further, an interaction protocol based on the affordances listed in the TDs).

We are exploring the former approach by introducing a novel way of machine-understandable description of \textit{coordination protocols} which enhances message specifications by embedding the semantics of the physical process, a concept that has not been previously explored.
We do this by extending the message descriptions with semantics about the \textit{physical} pre- and post-conditions (both on the part of the caller and the receiver). 
For example, the pre- and post-conditions referring to physical states can be obtained from the state machine models of the control programs of the agents.
An ontology provides this formalization, which we make available on our project website.

Further, instead of describing low-level protocol details and message schemes directly as a part of the coordination protocol, we include only the \emph{semantic type of the affordance} to search for in the TD, which, on the other hand, is more suited for conveying network and application protocol bindings.

Such coordination protocols hold potential by allowing us to model well-known interaction scenarios in various domains while carrying \textit{in-band knowledge} about the physical requirements and consequences of the interactions. For instance, they can be used to create demand-response protocols in energy management and be made readily available as Web resources in the hypermedia environment.
The following code snippet shows how a hypothetical message for flow change can be sent by a chiller whose valves are open. 
%The pointer to the TD affordance makes the specficication loosely-coupled to implementation specifics.

\begin{lstlisting}[language=SPARQL]
:request-flowrate-change a intr:Message.
 intr:hasSender :_sender.
:_sender a hvac:ChillerManager. 
 intr:precondition :_valves_open;
 intr:desiredPostcondition :_has_input_flow.
intr:hasReceiver :_reciever.
:receiver a hvac:PumpManager
 intr:precondition :_is_primed;
 intr:commitment :_flow_changed.
 intr:hasAffordance :_td_af_flowchange.
:_td_action_flowchange a hvac:FlowModulation.
\end{lstlisting}

\section{Discussion and Conclusion}
We have shown how machine-understandable descriptions of the engineering system, inclusive of the model of physical processes, enable automated inference of the agents' coordination responsibilities. While seeking to tackle the challenge of dynamically enacting coordination without requiring manual engineering, we see strong conceptual synergies in hypermedia-driven interactions. Hypermedia environments consisting of knowledge graphs, resources described in a machine-understandable manner, and support for large-scale deployments provide ideal support for automation agents operating in a distributed and dynamic environment.
Our research efforts in the past decade have laid the foundations by enabling the creation of integrated system descriptions and infusing knowledge of physics into digital representations.

However, our work is far from complete. There is a pressing need for further research to delve into topics such as modeling coordination plans (e.g., to include temporal logic), norms and constraints, and the crucial aspect of handling failures.
By showing how hypermedia-based multi-agent systems could support future methods addressing adaptive automation, we hope to motivate researchers %towards a paradigm shift from the 
to rethink architectural paradigms which result in tight coupling.

\bibliographystyle{splncs04}
\bibliography{coordination}

\end{document}